\documentclass[aps,twocolumn,showpacs,fleqn]{revtex4}

\usepackage{graphicx}
\usepackage{amsmath}
\usepackage{amsfonts}
\usepackage{amssymb}
\usepackage{epstopdf}

\newcommand{\mean}[1]{\left\langle #1 \right\rangle}

\newcommand{\Tr}{\text{Tr}\,}
\renewcommand{\Im}{\text{Im}\,}
\renewcommand{\Re}{\text{Re}\,}
\renewcommand{\imath}{\mathrm{i}}

\newcommand{\boldgreek}[1]{\ensuremath{\mbox{\boldmath$#1$}}}
\newcommand{\bg}[1]{\ensuremath{\mbox{\boldmath$#1$}}}

\newcommand{\RDM}{\ensuremath{\sigma}}			

\newcommand{\COMM}[2]{\left[#1, #2 \right]_-}	

\begin{document}
\title{Propagation Scheme for Non-Equilibrium Dynamics of Electron Transport
  in Nanoscale Devices}
\author{Alexander Croy}
\email{croy@pks.mpg.de}

\author{Ulf Saalmann}
	
\affiliation{Max-Planck-Institute for the Physics of Complex Systems,
  N\"{o}thnitzer Str.~38, 01187 Dresden, Germany}
	
\date{\today}

\begin{abstract}\noindent
  A closed set of coupled equations of motion for the description of
  time-dependent electron transport is derived. It provides the time
  evolution of energy-resolved quantities constructed from non-equilibrium
  Green functions. By means of an auxiliary-mode expansion a viable
  propagation scheme for finite temperatures is obtained, which allows to 
  study arbitrary time dependences and structured reservoirs.
  Two illustrative examples are presented.
\end{abstract}
	
	
\maketitle
\section{Introduction}\label{sec:Intro}
The investigation of time-resolved currents in mesoscopic devices has
gained a lot of interest over the past few years. This is not only because
of the potential application to quantum computing but also due to the
advent of new experiments specifically looking into time-dependent
electron transport \cite{fuha+06,koel+06}.
For example, manipulation of quantum dot systems is performed by using 
pump-probe schemes with a single voltage pulse.
The rising and the falling edge of the pulse lead to pumping 
and probing the device, respectively.
The experiments include transient-current spectroscopy of single quantum-dots \cite{fuau+03} 
and coherent manipulation of charge \cite{hafu+03} and spin
\cite{pejo+05,kobu+06} qubits in double quantum dots (DQDs).

The theoretical description of the electric current through a device coupled to two electron reservoirs
is usually based on Keldysh non-equilibrium Green function 
(NEGF) techniques \cite{jawi+94,haja07}. Within this approach the description of piecewise constant 
and sinusoidal voltage pulses is readily possible in the wide-band limit (WBL).
For harmonic modulations more sophisticated methods
\cite{stwi96,cale+03,ar05} combining Floquet theory and NEGF
formalism have been developed and allow going beyond WBL.
In order to overcome the limitations of the special form of driving and of the WBL, schemes based 
on the traditional approach \cite{jawi+94}, but working directly in the time domain have been put forward 
\cite{zhma+05,mogu+07}. All time convolutions, which result from the projection onto the device
states, are transformed into matrix-matrix multiplications using a time-discretization scheme.
In contrast, the formalism presented in Refs.\ \cite{kust+05,stpe+08} is based on propagating 
the wave function of the full system (device and reservoirs). This is accomplished by using the Cayley propagator 
and then projecting on the subsystem of interest. This formalism provides a natural way to work within
the so-called partition-free approach, where the time-dependent voltage pulse
is considered to act on the total system.

In this article we present a general propagation scheme which is also based on the 
NEGF formalism \cite{jawi+94} and allows to obtain device-related observables 
and the electric current as a function of time. Our formulation, however, relies 
on a set of coupled equations of motion for quantities with only \emph{one\/}
time argument. In this way we can avoid time convolutions and standard
methods for integrating the equations may be applied. As with the standard
formulation, the key issue consists in performing integrals over
reservoir states which eventually lead to tunneling self-energies. In this
context we propose using an auxiliary-mode expansion which allows to
treat finite temperatures.
We provide a numerical implementation of our scheme for two relevant cases\,---\,
the wide-band limit and a level-width function given by a sum of Lorentzians. 
The methods are applied to transport through a randomly fluctuating level and the 
transient response of a DQD to a voltage pulse.

In the remaining part of the introduction we briefly discuss the general setup
[Sec.\ \ref{sec:IntroSetup}] and then repeat the findings of the standard NEGF
formalism in the context of our propagation scheme [Sec.\ \ref{sec:NEGFIntro}].

\subsection{Setup}\label{sec:IntroSetup}
We take the usual threefold setup consisting of a device (system) which is
coupled to two electron reservoirs. The coupling is due to tunneling through
a barrier. The total Hamiltonian is
\begin{equation}
  H = H_\mathrm{D} + H_\mathrm{R} + H_\mathrm{DR}\,.
  \label{eq:totalHam}
\end{equation}
The device is described in terms of discrete energy levels $\varepsilon_n(t)$ 
which may be coupled through $V_{n m}(t)$,
\begin{equation}\label{eq:DevHamilOp}
  H_\mathrm{D} = \sum_n \varepsilon_n(t) c^\dagger_n c_n
  +\sum_{n\ne m} V_{n m}(t) c^\dagger_n c_m \;.
\end{equation}
The operators $\{c^\dagger_n\}$ and $\{c_n\}$ denote the creation and annihilation 
of an electron in state $n$.
The reservoirs are described by non-interacting electrons and the Hamiltonian reads
\begin{equation}
  H_\mathrm{R} = \sum_{\alpha \in \mathrm{L},\mathrm{R}} 
  \sum_k \varepsilon_{\alpha k}(t) b^\dagger_{\alpha k} b_{\alpha k}\;
  \label{eq:ResHamilOp}
\end{equation}
with single-particle energies of the form 
$\varepsilon_{\alpha k}(t) = \varepsilon^0_{\alpha k} + \Delta_{\alpha k}(t)$.
Finally, the coupling Hamiltonian is
\begin{eqnarray}
  H_\mathrm{DR} &=& \sum_n \sum_{\alpha k }
      T^\alpha_{k n}(t) \,b^\dagger_{\alpha k} c_n
    + \rm{h.c.} \;,
 \label{eq:TunnHam}
\end{eqnarray}
with $\{T^\alpha_{k n}\}$ denoting the couplings between device and
reservoir $\alpha={\rm L}, {\rm R}$; $\{b^\dagger_{\alpha k}\}$ and $\{b_{\alpha k}\}$ are
electron creation and annihilation operators for reservoir
states, respectively.

Regarding the time dependence of the reservoirs and the device we
adopt the Caroli partition scheme \cite{caco+71}, 
i.e.\, all sub-systems are separated at $t=-\infty$
and in their respective equilibrium state. Any time dependence only
sets in after eventually coupling the different parts. Consequently, the
single-particle occupation probability in the reservoirs is determined by
$\varepsilon^0_{\alpha k}$; the time dependence $\Delta_{\alpha k}(t)$ of the reservoir energies
appears as a phase-factor only. 
The situation where the chemical potentials and therefore the occupation probabilities are time-dependent has been critically discussed before \cite{jawi+94}.

\subsection{Time-Dependent Current and Non-Equilibrium Green Functions}\label{sec:NEGFIntro}
By applying the Keldysh formalism to non-equilibrium Green
functions it is possible to obtain a general formula for the 
time-dependent current in the setup introduced in
Sec.\ \ref{sec:IntroSetup}.
The current $J_\alpha$ through the barrier connecting 
lead $\alpha$ and the device is given by \cite{jawi+94,haja07}
\begin{eqnarray}
  J_\alpha (t) 
  &=& 2 e\, \Re \Tr \left\{ \int\limits_{-\infty}^{\infty} dt_1 \left[ 
      \mathbf{G}^< (t, t_1) \boldgreek{\Sigma}^\mathrm{a}_\alpha (t_1,t)  						\right.\right. \notag\\
  &&\qquad\qquad \left.\left.+\mathbf{G}^\mathrm{r} (t, t_1) \boldgreek{\Sigma}^<_\alpha (t_1,t) \right] \vphantom{\int\limits_{-\infty}^{\infty}}\right\} \;.
\label{eq:GenCurrent1}
\end{eqnarray}
Throughout the paper we adopt units with $\hbar = 1$.
In Eq.\ \eqref{eq:GenCurrent1} $\mathbf{G}^<$ and $\mathbf{G}^\mathrm{r}$ are lesser and
retarded Green functions and $\boldgreek{\Sigma}^\mathrm{a}$ and
$\boldgreek{\Sigma}^<$ are advanced and lesser self-energies, respectively. 
All boldface quantities are matrices related to the device states,
e.g., $\mathbf{G}^< (t, t_1)\equiv G_{nm}^< (t, t_1)$.
Products are to be understood as matrix multiplications.
The greater and lesser self-energies are explicitly given by
\begin{subequations}
\label{eq:defself}
\begin{eqnarray}
	\boldgreek{\Sigma}^>_\alpha (t_1,t) &=& -\imath \int
        \frac{d\varepsilon}{2 \pi} \bar{f}_\alpha (\varepsilon)
        e^{-\imath \varepsilon (t_1 - t)}
        \boldgreek{\Gamma}_\alpha (\varepsilon, t_1, t)\;,
        \label{eq:grself}\\ 
	\boldgreek{\Sigma}^<_\alpha (t_1,t) &=& \imath \int
        \frac{d\varepsilon}{2 \pi} f_\alpha (\varepsilon)
        e^{-\imath \varepsilon (t_1 - t)}
        \boldgreek{\Gamma}_\alpha (\varepsilon, t_1, t) \;. 
	\label{eq:leself}
\end{eqnarray}
\end{subequations}
As indicated at the end of the previous section the Fermi distribution, 
$f_\alpha(\varepsilon)\equiv f( \beta (\varepsilon-\mu_\alpha) )$, characterizes the 
equilibrium state of reservoir $\alpha$ with the chemical
potential $\mu_\alpha$ and inverse temperature $\beta = (k_{\rm B} T)^{-1}$ 
at $t_0=-\infty$. 
It is $ \bar{f}_\alpha (\varepsilon)=1-f_\alpha (\varepsilon)$.
Using a relation for two-time functions,
\begin{equation}
  X^{\mathrm{r},\mathrm{a}}(t,t') = \pm \Theta(\pm t \mp t') \left[ X^>(t,t') - X^<(t,t') \right] \,,
  \label{eq:retadvgreles}
\end{equation}
which applies to Green functions as well as to self energies,
one can find advanced (retarded) self-energies in 
Eq.\ \eqref{eq:GenCurrent1} in terms of greater and lesser functions.
The level-width function $\boldgreek{\Gamma}_\alpha$ in Eqs.\ (\ref{eq:defself}) 
depends on the density of states $\rho_\alpha(\varepsilon)$ of reservoir $\alpha$ and the coupling 
$T_{\alpha, n}(\varepsilon)$ of device level $n$ and the reservoir state at 
energy $\varepsilon$,
\begin{align}
  \left[ \boldgreek{\Gamma}_\alpha (\varepsilon, t_1, t) \right]_{m n} =&  2 \pi \rho_\alpha (\varepsilon) 
  T_{\alpha, n}(\varepsilon, t) T^*_{\alpha, m}(\varepsilon, t_1) \notag\\
  &\times\exp\{\imath \int\limits_{t_1}^{t} dt_2 \Delta_\alpha (\varepsilon,t_2)\}\;.
  \label{eq:LevelWidth}
\end{align}
Replacing the advanced and retarded quantities in Eq.\ \eqref{eq:GenCurrent1}
by using Eq.\ \eqref{eq:retadvgreles} one can rewrite the expression for
the current in a very compact form,
\begin{equation}\label{eq:CurrPi}
  J_\alpha(t) = {2 e}\, \Re \Tr \left\{ \boldgreek{\Pi}_\alpha (t) \right\}\;.
\end{equation}
The {\it current matrices} $\bg{\Pi}_\alpha ( t )$ are given by the following expression
\begin{equation}
  \bg{\Pi}_\alpha ( t ) = \int\limits^t_{t_0} dt_2 
     \left( \mathbf{G}^>(t,t_2) \bg{\Sigma}^<_\alpha (t_2,t)
       - \mathbf{G}^<(t,t_2) \bg{\Sigma}^>_\alpha (t_2,t) \right) \;,
  \label{eq:DefPi}
\end{equation}
where the first and the second term describe electrons tunneling {\it into} 
and {\it out of} the device, respectively. Equations \eqref{eq:CurrPi} and
\eqref{eq:DefPi} have been discussed in the context of current conserving
self-energies \cite{haja07}. The conceptually new approach presented in
this article consists in considering $\bg{\Pi}_\alpha ( t )$ as an independent
entity. In particular, opposed to correlation functions such as $\mathbf{G}^{\gtrless}(t,t_2)$ the
current matrices $\bg{\Pi}_\alpha(t)$ only depend on a single time argument.
Therefore, they are fully determined by a single equation of motion.
This circumstance provides the basis of our propagation scheme,
which is presented in Sec.\ \ref{sec:EOM}.

Moreover, in order to calculate the expectation value of any device observable
$O_{\rm D}$ it is advantageous to use the reduced single-electron density matrix, 
$\bg{\sigma}(t) = \Im \mathbf{G}^{<}(t,t)$. The expectation value is then given by
\begin{equation}
  \mean{O_{\rm D}(t)} = \Tr_{\rm D} \left\{ \bg{O}_{\rm D} \bg{\sigma}(t) \right\}\;.
\end{equation}
Similar to the current matrices $\bg{\Pi}_\alpha(t)$ the density matrix only depends 
on a single time argument and one has the following equation of motion
\begin{eqnarray}
  \imath \frac{\partial}{\partial t}\bg{\RDM}( t )
    &=& \COMM{ \mathbf{H} (t) }{ \bg{\RDM}( t ) } 
    \nonumber\\ &&
    +\imath \sum\limits_\alpha \left( \bg{\Pi}_\alpha ( t ) + \bg{\Pi}^\dagger_\alpha ( t ) \right)\;,
    \label{eq:EOMSEDM1}
\end{eqnarray}
which depends on the current matrices $\bg{\Pi}_\alpha(t)$. The boldface Hamiltonian
$\mathbf{H} (t) \equiv \left[ H_{\rm D} \right]_{nm}$ is obtained from the device Hamiltonian
in Eq.\ \eqref{eq:DevHamilOp}. Equation \eqref{eq:EOMSEDM1} is found by 
using $\mathbf{G}^{<}(t',t) = -\left[\mathbf{G}^{<}(t,t')\right]^\dagger$ and from the equations of motion for greater 
and lesser Green functions $\mathbf{G}^>$ and $\mathbf{G}^<$,
\begin{eqnarray}
  \imath \frac{\partial}{\partial t}\mathbf{G}^{\gtrless}(t,t') 
  &=& \mathbf{H} (t) \mathbf{G}^{\gtrless}(t,t') \notag\\
   &&+ \int dt_2 \boldgreek{\Sigma}^\mathrm{r}_{\rm tot}(t,t_2) \mathbf{G}^{\gtrless}(t_2,t') \notag\\
   &&+ \int dt_2 \boldgreek{\Sigma}^{\gtrless}_{\rm tot}(t,t_2) \mathbf{G}^\mathrm{a}(t_2,t') \;.
   \label{eq:GlGgEOM}
\end{eqnarray}
The total self-energies $\boldgreek{\Sigma}^{\gtrless,\mathrm{r}}_{\rm tot}$ are sums of the
tunneling self-energies for each reservoir. The Green functions may also be obtained 
from the Dyson series leading to an integral equation \cite{haja07}.
	
\section{Current Matrices and Auxiliary Mode Expansion}\label{sec:EOM}
In order to arrive at a viable propagation scheme we will
rewrite the equations of motion given above by introducing
energy-resolved quantities.
This form allows for applying an auxiliary-mode expansion
which replaces the energy integrals by finite sums.
The number of (matrix) equations to be propagated is determined
by the size of the expansion.
\subsection{Energy-Resolved Current Matrices}
First we assume factorizing momentum and time dependence of the tunnel coupling, 
$T_{\alpha, n}(\varepsilon, t) = T_{\alpha, n}(\varepsilon) u_{\alpha, n}(t)$.
The same ansatz has been used in Ref.\ \cite{jawi+94} for the non-interacting
resonant-level model.
For notational convenience we consider in the following only the case
of a common time dependence of the coupling for all device
states, i.e.\, $u_{\alpha, n}(t) = u_{\alpha}(t)$. Equation \eqref{eq:LevelWidth} becomes
\begin{equation}
   \boldgreek{\Gamma}_\alpha (\varepsilon, t_1, t) 
   = u^*_\alpha(t_1) u_\alpha(t) \boldgreek{\Gamma}_\alpha (\varepsilon) 
     \exp\{\imath \int\limits_{t_1}^{t} dt_2 \Delta_\alpha (\varepsilon,t_2)\}\;.
   \label{eq:AssFacMomTim}
\end{equation}
Next, we define {\it energy-resolved self-energies} as
\begin{subequations}
\label{eq:defenself}
\begin{eqnarray}
  \boldgreek{\Sigma}^>_\alpha (\varepsilon; t_1,t) 
  &=& -\imath u^*_\alpha(t_1) \bar{f}_\alpha (\varepsilon) e^{-\imath \varepsilon (t_1 - t)}
        \boldgreek{\Gamma}_\alpha (\varepsilon) \notag \\
   &&   \quad\times \exp\{\imath \int\limits_{t_1}^{t} dt_2 \Delta_\alpha (\varepsilon,t_2)\}\;,\label{eq:grreself}\\ 
  \boldgreek{\Sigma}^<_\alpha (\varepsilon; t_1,t) 
  &=& \imath u^*_\alpha(t_1) f_\alpha (\varepsilon) e^{-\imath \varepsilon (t_1 - t)}
        \boldgreek{\Gamma}_\alpha (\varepsilon) \notag \\
   &&   \quad\times  \exp\{\imath \int\limits_{t_1}^{t} dt_2 \Delta_\alpha (\varepsilon,t_2)\}\;.
\label{eq:lereself}
\end{eqnarray}
\end{subequations}
In terms of these expressions the full self-energies are given by
\begin{eqnarray}
  \boldgreek{\Sigma}^{\gtrless}_\alpha ( t_1,t) 
  &=& u_\alpha(t) \int d\varepsilon 
      \boldgreek{\Sigma}^{\gtrless}_\alpha (\varepsilon; t_1, t) \;,
      \label{eq:fullself}
\end{eqnarray}
which follows from Eq.\ \eqref{eq:defself}.
Using the definitions above we introduce {\it energy-resolved current matrices},
\begin{eqnarray}
  \bg{\Pi}_\alpha (\varepsilon; t) =&& \int\limits^t_{t_0} dt_2  
     \left( \mathbf{G}^>(t,t_2) \bg{\Sigma}^<_\alpha (\varepsilon; t_2,t) \right. \notag\\
     &&\quad\left.
       - \mathbf{G}^<(t,t_2) \bg{\Sigma}^>_\alpha (\varepsilon; t_2,t) \right) \;.
  \label{eq:EresPi}
\end{eqnarray}
From Eq.\ \eqref{eq:EresPi} one finds $ \bg{\Pi}_\alpha (\varepsilon; t_0) = \bg{0}$.
The expression for the current given by Eq.\ \eqref{eq:CurrPi} becomes
\begin{equation}\label{eq:CurrPiEres}
  J_\alpha(t) = {2 e}\, \Re \sum\limits_n u_\alpha(t) \int d\varepsilon 
      \Pi_{\alpha, nn} ( \varepsilon; t) \;.
\end{equation}
Therefore, the diagonal elements $\Pi_{\alpha, nn} ( \varepsilon; t)$ may be interpreted
as the current flowing from the reservoir state at energy $\varepsilon$ to the system
state $n$. The total current through the barrier is then given by the
sum of all possible currents.

The equation of motion [Eq.\ \eqref{eq:EOMSEDM1}] for the reduced single-electron density matrix 
$\bg{\sigma}$ of the device becomes
\begin{eqnarray}\label{eq:EOMSEDM2}
  \imath \frac{\partial}{\partial t}\bg{\sigma}(t)
  &=& \COMM{ \mathbf{H} (t) }{ \bg{\sigma}(t)  } \\
   && +\imath \sum\limits_\alpha \int d\varepsilon 
    \left( u_\alpha(t) \bg{\Pi}_\alpha ( \varepsilon; t ) 
         + u^*_\alpha(t) \bg{\Pi}^\dagger_\alpha ( \varepsilon; t ) 
    \right) \;, \notag
\end{eqnarray}
which now contains the energy-resolved current matrices.

Due to the definitions (\ref{eq:defenself}) of the
energy-resolved self-energies, their time derivatives,
\begin{equation}
  \frac{\partial}{\partial t} \boldgreek{\Sigma}^{\gtrless}_\alpha (\varepsilon; t_1,t) 
  = \imath \left( \varepsilon + \Delta_\alpha (\varepsilon,t) \right)
    \boldgreek{\Sigma}^{\gtrless}_\alpha (\varepsilon; t_1,t) \;,
  \label{eq:EresSelfEOM}
\end{equation}
and by using Eq.\ \eqref{eq:GlGgEOM}, one gets an equation of motion for the 
energy-resolved current matrices,
\begin{eqnarray}
  \imath \frac{\partial}{\partial t} \bg{\Pi}_\alpha (\varepsilon; t)
  &=& -\frac{\imath}{2 \pi} u^*_\alpha (t) 
      \left( 
        \bg{\sigma}(t) - f_\alpha (\varepsilon) 
      \right) \bg{\Gamma}_\alpha (\varepsilon) \notag\\
   && + \left\{ 
     \mathbf{H}(t) - \left( \varepsilon + \Delta_\alpha (\varepsilon,t) \right) 
      \right\} \bg{\Pi}_\alpha (\varepsilon; t)  \notag\\
   && + \sum\limits_{\alpha'} u^*_{\alpha'}(t) \int
   d\varepsilon'\, \bg{\Omega}_{\alpha \alpha'}(\varepsilon,
   \varepsilon'; t) \;, \notag\\ &&\label{eq:EOMPi}
\end{eqnarray}
where a new quantity $\bg{\Omega}_{\alpha \alpha'}$ has to be introduced. 
It contains all contributions from the time derivative of the greater and
lesser Green functions, which give rise to a double time integral.
Consequently, its definition is 
\begin{widetext}
\begin{eqnarray}\label{eq:DefOmega}
  \bg{\Omega}_{\alpha \alpha'}(\varepsilon, \varepsilon'; t) 
  &=& \int\limits^t_{t_0} dt_2  \int\limits^t_{t_0} dt_1 
      \bg{\Sigma}^r_{\alpha'}(\varepsilon'; t,t_1)
      \left[ 
         \mathbf{G}^>(t_1,t_2) \bg{\Sigma}^<_\alpha(\varepsilon; t_2,t)
        - \mathbf{G}^<(t_1,t_2) \bg{\Sigma}^>_\alpha(\varepsilon; t_2,t)        
      \right] \\
  && - \int\limits^t_{t_0} dt_2  \int\limits^{t_2}_{t_0} dt_1 
      \left[ 
        \bg{\Sigma}^<_{\alpha'}(\varepsilon'; t,t_1) \mathbf{G}^a(t_1,t_2) \bg{\Sigma}^>_\alpha(\varepsilon; t_2,t)
        - \bg{\Sigma}^>_{\alpha'}(\varepsilon'; t,t_1) \mathbf{G}^a(t_1,t_2) \bg{\Sigma}^<_\alpha(\varepsilon; t_2,t)
      \right] \;. \notag
\end{eqnarray}
We replace the retarded self-energies and the advanced Green function again using
Eq.\ \eqref{eq:retadvgreles}, but instead of showing the result we rather
give the equation of motion, which is easily obtained from Eq.\ \eqref{eq:DefOmega},
\begin{eqnarray}
  \imath \frac{\partial}{\partial t} \bg{\Omega}_{\alpha \alpha'}(\varepsilon, \varepsilon'; t) 
  &=& \frac{1}{2\pi} \left\{
  u_{\alpha'}(t) \bg{\Gamma}_{\alpha'} (\varepsilon') \bg{\Pi}_\alpha (\varepsilon; t)
  + \bg{\Pi}^\dagger_{\alpha'} (\varepsilon'; t) \bg{\Gamma}_\alpha (\varepsilon) u^*_{\alpha}(t)\right\}
\nonumber\\
  &&+ \big\{ 
    \left( \varepsilon' + \Delta_{\alpha'} (\varepsilon',t) \right)
    - \left( \varepsilon + \Delta_\alpha (\varepsilon,t) \right)
  \big\}\:\bg{\Omega}_{\alpha \alpha'}(\varepsilon, \varepsilon'; t) \;,
  \label{eq:EOMOmega}
\end{eqnarray}
\end{widetext}
with the initial conditions $\bg{\Omega}_{\alpha \alpha'}(\varepsilon, \varepsilon'; t_0) = \bg{0}$.
The equations of motion given by Eqs.\ \eqref{eq:EOMSEDM2}, \eqref{eq:EOMPi} 
and \eqref{eq:EOMOmega} provide a closed description of the non-equilibrium dynamics
of the device. A similar set of equations has been found recently \cite{jizh+08}, 
where it was derived from a hierarchy for the many-body density matrix. 
The identification of
\begin{equation}
  \bg{\varphi}_\alpha = -\imath \bg{\Pi}_\alpha \quad\text{and}\quad
  \bg{\varphi}_{\alpha' \alpha} = -\imath \bg{\Omega}_{\alpha \alpha'}
\end{equation}
renders their equations identical to the ones given above. This provides an independent 
verification of the density-matrix approach \cite{jizh+08} and shows that the 
hierarchy derived therein yields the exact dynamics for non-interacting electrons under 
the assumptions stated above.

The full single-particle density matrix has a size of $(N_{\rm D} + N_{\rm R})^2$, where $N_{\rm D}$
and $N_{\rm R}$ are the number of single-particle states in the device and the reservoirs, respectively. 
In the present case we have to propagate $N_{\rm D}^2 \times (N_{\rm R} + 1)^2$ quantities with $\bg{\Pi}$ and $\bg{\Pi}^\dagger$
counting independently.
Therefore, the complexity of Eqs.\ \eqref{eq:EOMSEDM2}, \eqref{eq:EOMPi} and \eqref{eq:EOMOmega} is at least 
the same compared to calculating the full single-particle density matrix. In particular one has to deal 
with a continuum of states and consequently the utility of the method depends on finding an efficient 
strategy for performing the energy integral. 
In the following subsection we will provide such a method based on the expansion of the Fermi function and making use of
the residue theorem. The same strategy has been successfully applied to the propagation of non-Markovian quantum master equations
involving bosonic \cite{meta99} and fermionic reservoirs \cite{wesc+06,jizh+08}.
The formulation in terms of energy-resolved quantities depending on a single time argument turns 
out to be beneficial in this context. In order to propagate each matrix only the value of
the previous time step has to be known. References to past times
\cite{zhma+05,mogu+07} are not necessary. This comes at the cost of having to propagate 
the two-energy quantity $\bg{\Omega}_{\alpha \alpha'}$. However, as we will show in the
next section one can effectively reduce the associated numerical costs by using
an auxiliary-mode expansion.

\subsection{Auxiliary-Mode Expansion} \label{sec:AuxModes}
The general idea of the auxiliary-mode expansion consists in making use of contour integration and
the residue theorem. To this end the Fermi function is expanded in a sum over $N_\mathrm{F}$ simple poles,
\begin{equation}
  f_\alpha(\varepsilon)
  \approx \frac{1}{2} - \frac{1}{\beta}\sum_{p=1}^{N_\mathrm{F}}\left(
    \frac{1}{\varepsilon{-}\chi_{\alpha p}^+}
    +\frac{1}{\varepsilon{-}\chi_{\alpha p}^-}\right)
  \label{eq:ExpFermiFun}
\end{equation}
with $\chi_{\alpha p}^\pm = \mu_\alpha{\pm}x_p/\beta$ and $\Im x_p >0$. The well-known
Matsubara expansion \cite{ma90} is an example for such a decomposition.
Its major disadvantage consists in a poor convergence behavior especially for
low temperatures. A particular efficient alternative is presented in 
appendix \ref{sec:AppExp}.

\subsubsection{Wide-Band Limit}\label{sec:WBLExp}
As a first application we consider the WBL, 
i.e.\, $\bg{\Gamma}_\alpha(\varepsilon)=\text{const}$.
From the definition of the self-energies \eqref{eq:defself} and 
the expansion of the Fermi function [Eq.\ \eqref{eq:ExpFermiFun}] one obtains
for $t>t_1$,
\begin{eqnarray}
  \boldgreek{\Sigma}^>_\alpha (t_1,t) 
  &=& -\imath \frac{1}{2} \bg{\Gamma}_\alpha |u_\alpha(t)|^2 \delta(t-t_1) \\
  && + u_\alpha(t) \sum\limits_p \frac{1}{\beta} \bg{\Gamma}_\alpha u^*_\alpha(t_1)
      e^{\imath \int^t_{t_1} dt_2 \chi^+_{\alpha p} (t_2) }\;,\notag
\end{eqnarray}
where $\chi^+_{\alpha p} (t) = \chi^+_{\alpha p} + \Delta_\alpha (t)$.
Analogously, one finds for the lesser self-energy
\begin{eqnarray}
  \boldgreek{\Sigma}^<_\alpha (t_1,t) 
  &=& \imath \frac{1}{2} \bg{\Gamma}_\alpha |u_\alpha(t)|^2 \delta(t-t_1)  \\
  && + u_\alpha(t) \sum\limits_p \frac{1}{\beta} \bg{\Gamma}_\alpha u^*_\alpha(t_1)
      e^{\imath \int^t_{t_1} dt_2 \chi^+_{\alpha p} (t_2)}\;. \notag
\end{eqnarray}
Thus, the expansion of the Fermi function leads to an expansion of the self-energies
into a sum of exponentials. Due to the WBL one also gets one term proportional to
a delta function.
We introduce {\it auxiliary self-energies} $\boldgreek{\Sigma}_{\alpha p}$,
which incorporate the exponentials, i.e.\
\begin{subequations}\label{eq:selfexp}
\begin{eqnarray}
  \boldgreek{\Sigma}^{\gtrless}_\alpha (t_1,t) 
  &=& \mp\imath \frac{1}{2} \bg{\Gamma}_\alpha |u_\alpha(t)|^2 \delta(t-t_1) \notag\\
  &&  + u_\alpha(t) \sum\limits_p \boldgreek{\Sigma}_{\alpha p} (t_1,t) \;, \\
\boldgreek{\Sigma}_{\alpha p} (t_1,t) &=&  \frac{1}{\beta} \bg{\Gamma}_\alpha u^*_\alpha(t_1)
 e^{\imath \int^t_{t_1} dt_2 \chi^+_{\alpha p} (t_2)}\;,
\end{eqnarray}
\end{subequations}
which implies $\boldgreek{\Sigma}_{\alpha p} (t,t_+) = \frac{1}{\beta} \bg{\Gamma}_\alpha u^*_\alpha(t)$.
Next, we insert the expanded self-energies into the definition of the current matrices 
\eqref{eq:DefPi},
\begin{eqnarray}  \bg{\Pi}_\alpha ( t ) 
  &=& \frac{1}{4} |u_\alpha(t)|^2 \left(  \mathbf{1} - 2 \bg{\sigma}(t) \right) 
      \bg{\Gamma}_\alpha \notag\\
  &&  \quad + u_\alpha(t) \sum\limits_p \bg{\Pi}_{\alpha p} ( t ) \;, 
\label{eq:DefPiWBL}
\end{eqnarray}
and obtain an expansion in terms of {\it auxiliary current matrices},
\begin{eqnarray}
  \bg{\Pi}_{\alpha p} ( t ) &=&
  \int\limits^t_{t_0} dt_2 
       \left( \mathbf{G}^<(t,t_2) \bg{\Sigma}_{\alpha p} (t_2,t) \right.\notag\\
  &&\left.\quad     - \mathbf{G}^>(t,t_2) \bg{\Sigma}_{\alpha p} (t_2,t) \right) \;. 
\end{eqnarray}
Their equation of motion is easily found,
\begin{eqnarray}\label{eq:PiAuxWBLEOM}
  \imath \frac{\partial}{\partial t} \bg{\Pi}_{\alpha p} ( t )
  &=& \frac{1}{\beta} \bg{\Gamma}_\alpha u^*_\alpha(t) \\
  && + \left( \mathbf{H}(t) -\frac{\imath}{2} \bg{\Gamma}(t) -\chi^+_{\alpha p} (t) \mathbf{1} \right)  \bg{\Pi}_{\alpha p} ( t )\;, \notag
\end{eqnarray}
where $\bg{\Gamma}(t) = \sum_{\alpha'} |u_{\alpha'}(t)|^2 \bg{\Gamma}_{\alpha'}$.
The coupled equations of motion \eqref{eq:EOMSEDM1} and \eqref{eq:PiAuxWBLEOM} 
allow with Eq.\ \eqref{eq:DefPiWBL}
for a complete description of the non-equilibrium dynamics of the device.
Comparing Eqs.\ \eqref{eq:EOMPi} and \eqref{eq:PiAuxWBLEOM} suggests that
$\bg{\Omega}_{\alpha p, \alpha' p'}(t) = -\frac{\imath}{2} u_{\alpha'}(t) \bg{\Gamma}_{\alpha'}
\bg{\Pi}_{\alpha p} ( t ) \delta_{p p'}$. Thus, an additional equation of motion
for $\bg{\Omega}$ is not needed for the WBL.

\subsubsection{Lorentzian Level-Width Function}\label{sec:LLWF}
The next application we consider is the case of a Lorentzian level-width function (LLWF). We take
a general ansatz of the form
\begin{equation}
  \bg{\Gamma}_\alpha (\varepsilon) 
  = \sum\limits^{N_{\rm L}}_{\ell=1} 
  \left( 
    \frac{\bg{\Gamma}^+_{\alpha \ell}}{\varepsilon - \varepsilon_{\alpha \ell} -\imath W_{\alpha \ell}}
    + \frac{\bg{\Gamma}^-_{\alpha \ell}}{\varepsilon - \varepsilon_{\alpha \ell} +\imath W_{\alpha \ell}}
  \right)\;,
  \label{eq:LWFexp}
\end{equation}
with $W_{\alpha \ell} > 0$ and $\bg{\Gamma}^\pm_{\alpha \ell} = \mp \frac{\imath}{2}  \bg{\Gamma}_{\alpha \ell} W_{\alpha \ell}$. 
Equation \eqref{eq:LWFexp} might be used as a parametrization of an arbitrary level-width function \cite{meta99,wesc+06}.
Now, we can plug Eq.\ \eqref{eq:LWFexp} into the definition of the self-energies [Eq.\ \eqref{eq:defself}] and evaluate
the energy integral by means of contour integration. This procedure yields for $t>t_1$
\begin{subequations}\label{eq:selflor}
\begin{eqnarray}
  \boldgreek{\Sigma}^{>}_\alpha (t_1,t)
  &=&
  + u^*_\alpha(t_1) u_\alpha(t)\left( \sum\limits_\ell \bg{\Gamma}^+_{\alpha \ell} 
    \bar{f}^{\rm P}_{\alpha \ell} e^{-\imath (\varepsilon_{\alpha \ell} +\imath W_{\alpha \ell}) (t_1 - t)} \right.\notag\\
   && + \left.\sum\limits_p \frac{1}{\beta} \bg{\Gamma}_{\alpha} (\chi^+_{\alpha p}) e^{-\imath \chi^+_{\alpha p} (t_1 - t)} \right) \notag\\
   &&   \quad\times \exp\{\imath \int\limits_{t_1}^{t} dt_2 \Delta_\alpha (\varepsilon,t_2)\}\;,\\
  \boldgreek{\Sigma}^{<}_\alpha (t_1,t) 
  &=& - u^*_\alpha(t_1) u_\alpha(t)\left( \sum\limits_\ell \bg{\Gamma}^+_{\alpha \ell} 
    f^{\rm P}_{\alpha \ell} e^{-\imath (\varepsilon_{\alpha \ell} +\imath W_{\alpha \ell}) (t_1 - t)} \right.\notag\\
   && - \left.\sum\limits_p \frac{1}{\beta} \bg{\Gamma}_{\alpha} (\chi^+_{\alpha p}) e^{-\imath \chi^+_{\alpha p} (t_1 - t)} \right) \notag\\
   &&   \quad\times \exp\{\imath \int\limits_{t_1}^{t} dt_2 \Delta_\alpha (\varepsilon,t_2)\}\;, 
\end{eqnarray}
\end{subequations}
where $f^{\rm P}_{\alpha \ell} = f_\alpha (\varepsilon_{\alpha \ell} +\imath W_{\alpha \ell})$ indicates 
that the expansion given in Eq.\ \eqref{eq:ExpFermiFun} should be
used to calculate the Fermi function at the position of the pole $\ell$. 
The self-energies are thus given by a finite sum with $N_{\rm L} + N_{\rm F}$ terms. For convenience we combine the two indices $p$ and $\ell$ yielding a single index $x = \{\ell, p\}$. The coefficients and exponents are combined in a similar way, 
\begin{subequations}\label{eq:coefflor}
\begin{eqnarray}
  \bg{\Gamma}^{>,\pm}_{\alpha x} &=& 
      \{ \pm \bg{\Gamma}^{\pm}_{\alpha \ell} \bar{f}_\alpha (\varepsilon_{\alpha \ell} \pm \imath W_{\alpha \ell}) ,  
         \pm \frac{1}{\beta} \bg{\Gamma}_{\alpha} (\chi^\pm_{\alpha p})\}\;,\\
  \bg{\Gamma}^{<,\pm}_{\alpha x} 
  &=& \{ \mp \bg{\Gamma}^{\pm}_{\alpha \ell} f_\alpha (\varepsilon_{\alpha \ell} \pm \imath W_{\alpha \ell}) ,  
         \pm \frac{1}{\beta} \bg{\Gamma}_{\alpha} (\chi^\pm_{\alpha p})\}\;,\\
  \chi^\pm_{\alpha x} &=& \{ \varepsilon_{\alpha \ell} \pm \imath W_{\alpha \ell}, \chi^\pm_{\alpha p} \}\;.
\end{eqnarray}
\end{subequations}
Using these conventions the self-energies can be written in a compact form, assuming $t>t_1$ we have
\begin{subequations}
  \begin{eqnarray}
    \boldgreek{\Sigma}^{\gtrless}_\alpha (t_1,t) &=&  u_\alpha(t) \sum\limits_x \boldgreek{\Sigma}^{\gtrless}_{ \alpha x} (t_1,t)\;, \\
    \boldgreek{\Sigma}^{\gtrless}_{ \alpha x} (t_1,t) &=&  u^*_\alpha(t_1) \bg{\Gamma}^{\gtrless, +}_{\alpha x} e^{\imath \int^t_{t_1} dt_2 \chi^+_{\alpha x} (t_2)}\;,
  \end{eqnarray}
\end{subequations}
where $\chi^\pm_{\alpha x} (t) = \chi^\pm_{\alpha x} + \Delta_\alpha (t)$. The {\it auxiliary self-energies}
$\boldgreek{\Sigma}^{\gtrless}_{ \alpha x}$ are simply exponentials.
The respective {\it auxiliary current matrices} can be calculated in analogy 
to the energy-resolved current matrices, i.e.\,
\begin{eqnarray}
  \bg{\Pi}_{\alpha x} ( t ) 
  &=&\int\limits^t_{t_0} dt_2 
       \left( \mathbf{G}^>(t,t_2) \bg{\Sigma}^<_{\alpha x} (t_2,t) \right. \notag\\
   &&\quad\left.
       - \mathbf{G}^<(t,t_2) \bg{\Sigma}^>_{\alpha x} (t_2,t) \right)\;.
     \label{eq:PiLorDef}
\end{eqnarray}
Their equation of motion is then given by
\begin{eqnarray} 
  \imath \frac{\partial}{\partial t} \bg{\Pi}_{\alpha x} ( t )
  &=&  u^*_\alpha(t) \bg{\Gamma}^{<,+}_{\alpha x}
       + u^*_\alpha(t) \bg{\sigma}(t) \left( 
         \bg{\Gamma}^{>,+}_{\alpha x} - \bg{\Gamma}^{<,+}_{\alpha x} \right) \notag\\
   &&+ \left( \mathbf{H}(t) - \chi^+_{\alpha x}(t) \right) \bg{\Pi}_{\alpha x} (t) \notag\\
   &&+ \sum\limits_{\alpha' x'} u^*_{\alpha'}(t) \bg{\Omega}_{\alpha x, \alpha' x'}(t)
  \;. 
\label{eq:EOMPiLor}
\end{eqnarray}
The initial condition $\bg{\Pi}_{\alpha x} ( t_0 ) = \bg{0}$ follows from Eq.\ \eqref{eq:PiLorDef}.
Notice the similarity to the energy-resolved current matrices given by 
Eq.\ \eqref{eq:EOMPi}. In particular, we also have a two-mode quantity 
$\bg{\Omega}_{\alpha x, \alpha' x'}$ appearing in the equation of motion.
Its definition is again in full analogy to the energy-resolved case given 
in Eq.\ \eqref{eq:DefOmega}, but with $\bg{\Sigma}^{\gtrless}_{\alpha'}(\varepsilon'; t,t_1)$
replaced by $\bg{\Sigma}^{\gtrless}_{\alpha' x'}(t,t_1)$.
Also the equation of motion is similar to the energy-resolved case [Eq.\ \eqref{eq:EOMOmega}],
\begin{eqnarray}\label{eq:EOMOmegaLor}
  \imath \frac{\partial}{\partial t} \bg{\Omega}_{\alpha x, \alpha' x'}(t) 
  &=& \imath u_{\alpha'}(t) \left( 
      \bg{\Gamma}^{>,-}_{\alpha' x'} - \bg{\Gamma}^{<,-}_{\alpha' x'}
    \right) \bg{\Pi}_{\alpha x}(t)  \\
   && + \imath \bg{\Pi}^\dagger_{\alpha' x'} (t) 
      \left( 
        \bg{\Gamma}^{>,+}_{\alpha x} - \bg{\Gamma}^{<,+}_{\alpha x}
      \right) u^*_\alpha(t) \notag \\
   && + \left( \chi^-_{\alpha' x'}(t) - \chi^+_{\alpha x}(t) \right) \bg{\Omega}_{\alpha x, \alpha' x'}(t) \;.
  \notag
\end{eqnarray}
At $t=t_0$ one finds $\bg{\Omega}_{\alpha x, \alpha' x'}(t_0) = \bg{0}$. It is interesting to 
notice that for $x=p, x'=p'$, i.e.\, both indices represent an auxiliary mode resulting from the
Fermi-function expansion [Eq.\ \eqref{eq:ExpFermiFun}], one gets 
$\frac{\partial}{\partial t}\bg{\Omega}_{\alpha p, \alpha' p'}(t) \propto \bg{\Omega}_{\alpha p, \alpha' p'}(t)$. 
Taking the initial condition into account it follows that
$\bg{\Omega}_{\alpha p, \alpha' p'}(t)\equiv\bg{0}$   
for all times. 
This is consistent with the energy-resolved expression [Eq.\ \eqref{eq:EOMOmega}] where
any reference to the Fermi function is absent. Consequently, instead of propagating 
$(N_{\rm L} + N_{\rm F}) \times (N_{\rm L} + N_{\rm F})$ $\bg{\Omega}$-matrices we only need to consider
$N_{\rm L} \times (N_{\rm L} + 2 N_{\rm F})$ matrices for each reservoir index $\alpha$.
Since typically $N_{\rm L} \ll N_{\rm F}$ the memory requirement of the proposed method scales 
with $N_{\rm L}\times N_{\rm F}$ and the computational time requirement scales with 
$N_{\rm T}\times N_{\rm L}\times N_{\rm F}$, where
$N_{\rm T}$ is the number of time steps. Notice that in spite of having to use two-energy
quantities, using the auxiliary-mode expansion for the Fermi function yields a scheme, which scales \emph{linearly\/} with the number of modes and thus allows for a particularly efficient propagation.

\section{Applications}\label{sec:Appl}
We apply the proposed propagation scheme to two situations:
a resonant-level model with a randomly fluctuating energy level
and a DQD system driven by finite bias-voltage pulses.
These two situations demonstrate that our scheme is especially 
suited to study a strongly fluctuating driving and realistic
experimental pulses including structured reservoirs.

\subsection{Fluctuating Energy Level}\label{sec:RLMNoise}
\begin{figure}[b]
  \includegraphics[scale=0.6]{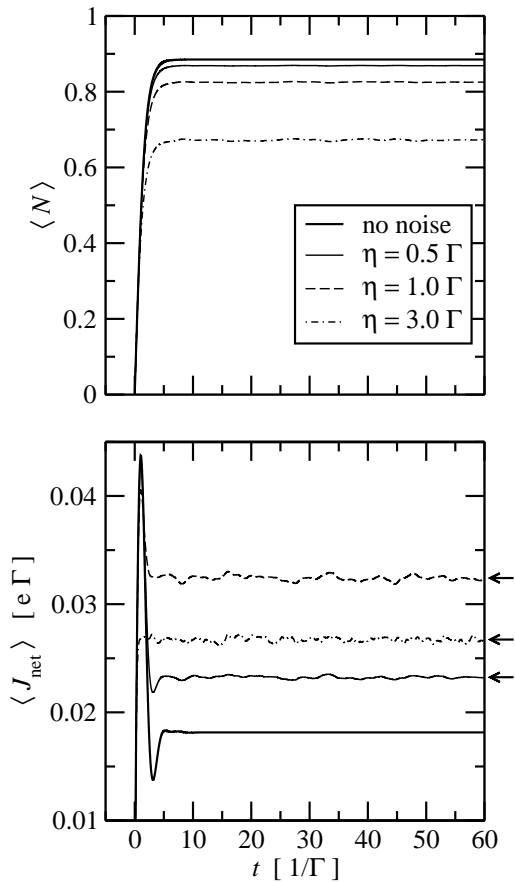}
  \caption{Time-resolved occupation $\mean{N}$ and net-current $\mean{J_{\rm net}}$ for different values of the noise amplitude. 
    Noise averages are obtained from $20000$ realizations and $\kappa = 0.5\Gamma$. The arrows 
    indicate the time-averaged values $\overline{ \mean{ J_{\rm net} }}$ obtained by sampling
    the current for times $t>10/\Gamma$.
	}\label{fig:AppRLMTimeCurr}   
\end{figure}%
As a first application we consider a resonant-level model with
a single randomly fluctuating energy-level $\varepsilon_{\rm d}(t)$,
which is given by a Gaussian stochastic process \footnote{The more general case of
coupling the spin degree of freedom of the tunneling electron to a 
fluctuating semi-magnetic barrier was investigated by 
Y.~G. Rubo, J. Exp. Theor. Phys. {\bf 77},  685  (1993).}.
An analytic expression for the current is given in appendix \ref{sec:AppNoise}.

The device Hamiltonian [Eq.\ \eqref{eq:DevHamilOp}] is simply,
\begin{equation}
        H_{\rm D} = \varepsilon_{\rm d} (t) \;c^\dagger_{\rm d} c_{\rm d}\;,
\end{equation}
with ``${\rm d}$'' denoting the single-electron state of the device. All matrices
become scalars and the respective equations of motion are scalar equations.
The stochastic process $\varepsilon_{\rm d} (t)$ is fully characterized
by the first and second moments,
\begin{subequations}\label{eq:NoiseProp}
\begin{eqnarray}
  \mean{ \varepsilon_{\rm d} (t) } &=& 0\;, \\
  \mean{ \varepsilon_{\rm d} (t) \varepsilon_{\rm d} (t') } &=& c(t - t')\;.
\end{eqnarray}
\end{subequations}
Here, we take $\varepsilon_{\rm d} (t)$ as realization of an
Ornstein-\/Uhlenbeck (OU) process, which yields for the correlation function 
$c(t-t') = \frac{\eta^2}{2 \kappa} \exp[-\kappa (t-t')]$. The OU process is characterized by
two parameters, the inverse correlation-time $\kappa$ and the noise amplitude $\eta$ \cite{ga96}.

Considering the WBL and using a symmetric coupling to the left and
right reservoir, $\Gamma_{\rm L} = \Gamma_{\rm R} = \Gamma/2$, 
we suddenly connect the device and the reservoirs at $t=0$.
The reservoirs are further characterized by chemical potentials $\mu_{\rm L} = 2\Gamma$,
$\mu_{\rm R}= \Gamma$ and temperature $k_{\rm B} T = 0.1\Gamma$. Thus, without stochastic
driving the energy level is not located in the transport window and a non-vanishing current is a result
of the broadening due to the coupling to the reservoirs.

The equations of motion obtained in Sect.\ \ref{sec:AuxModes} are propagated using a weak 
second-order Runge-Kutta scheme \cite{mitr94} with a constant time step
\footnote{For one set of parameters, $\kappa=0.25\Gamma$ and $\eta=5.0\Gamma$, we had to use $\delta t = 0.005/\Gamma$.}
$\delta t = 0.01/\Gamma$. We use
$N_{\rm F} = 240$ auxiliary modes for all calculations.
The resulting time-resolved occupation, $N(t)$, and net current,
$J_{\rm net}(t) = \left[J_{\rm L}(t) - J_{\rm R}(t)\right]/2$, 
are averaged over $20000$ realizations of the stochastic process. 
Figure \ref{fig:AppRLMTimeCurr} shows the averages $\left<N(t)\right>$ and $\left<J_{\rm net}(t)\right>$ for $\kappa=0.5\Gamma$ and three selected values of $\eta=0.5, 1.0, 3.0 \,\Gamma$. 
We also show the case of no stochastic driving. One sees a
transient response to the sudden coupling for times
$t=0\ldots10/\Gamma$ and 
the eventual settling to a stationary value. 
In all cases shown in Fig.\ \ref{fig:AppRLMTimeCurr} the
stationary current is larger than for the  
case without any noise; but the dependence on $\eta$ is
non-monotonic. 

\begin{figure}[htbp!]
  \includegraphics[scale=0.6]{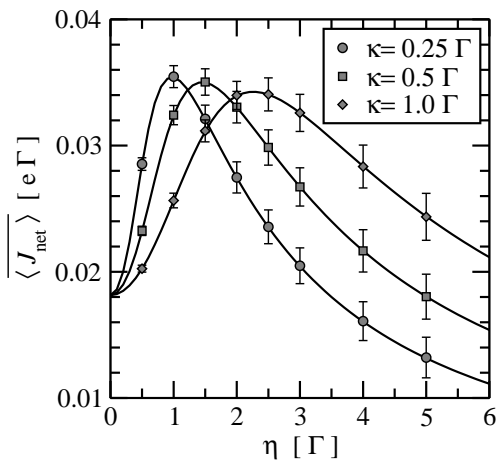}
  \caption{Time-averaged current vs noise amplitude for different values of the inverse noise correlation-time $\kappa$. 
    Noise averages are obtained   from $20000$ realizations. Error bars indicate $95\%$ confidence interval
    for the sample mean. Full lines denote the analytical result given by Eq.\ \eqref{eq:TaNAvgCurr}.}
  \label{fig:AppRLMStatCurr}   
\end{figure}

In order to quantify the stationary current we take the
time average $\overline{ \mean{ J_{\rm net} }}$ for the time interval 
starting at $t=10/\Gamma$. 
Figure \ref{fig:AppRLMStatCurr} shows the obtained time-averaged current
as a function of the noise strength $\eta$ and for various values of the inverse correlation-time $\kappa$.
The time-averaged current exhibits a pronounced maximum as a function of noise strength; the transport
through the energy-level is stochastically enhanced. This effect reminds of the phenomenon of stochastic
resonance \cite{gaha+98}. The observed behavior is a result of additional 
broadening due to the stochastic driving \cite{haja07}. The current is proportional to the area under the spectral 
density, $A(\varepsilon) = -2\Im G^r(\varepsilon)$, within the transport window given by $\{\mu_{\rm L}, \mu_{\rm R}\}$,
cf.\ appendix \ref{sec:AppNoise}.
For increasing noise strength the spectral density becomes broader and has more weight in the transport window. 
However, the height of $A(\varepsilon)$ decreases at the same time which eventually leads to a decrease in the 
area in the transport window. These findings are corroborated by the analytical result [Eq.\ \eqref{eq:TaNAvgCurr}], 
which is also shown in Fig.\ \ref{fig:AppRLMStatCurr}. The numerical results agree very well with those results. 

\subsection{Double Quantum Dot}
As a second application we will now discuss the response of a DQD to a voltage pulse. 
The device consists of two QDs which are coupled in series. Each dot is also coupled to an electron reservoir.
This setup resembles a typical experimental situation (see for example \cite{fuha+06}).

The DQD is modeled by a two-level system, i.e., one localized energy-level per dot. The device Hamiltonian
[Eq.\ \eqref{eq:DevHamilOp}] is then,
\begin{equation}
        H_{\rm D} = \sum\limits_{d = {\rm l,r}} \varepsilon_{d} (t) \;c^\dagger_{d} c_{d}
                   + V  c^\dagger_{\rm l} c_{\rm r} + {\rm h.c.}\;,
\end{equation}
with ``${\rm l}$'' and ``${\rm r}$'' the localized single-electron states. The time-dependent bias-voltage
is assumed to act on the energies in the following way: $\Delta_{\rm L} (t) = -\Delta_{\rm R} (t) = V_{\rm bias}(t)/2$ and
$\varepsilon_{\rm l} (t) = -\varepsilon_{\rm r} (t) = V_{\rm bias}(t)/4$. Initially, the chemical potentials
$\mu_{\rm L}$ and $\mu_{\rm R}$ and the QD energies $\varepsilon_{\rm l,r}$ are zero. The
temperature is $k_{\rm B} T = 0.1 \Gamma$ for both reservoirs.
Since the two dots are coupled in series, the level-width functions contain one non-zero element,
\begin{gather*}
  \boldgreek{\Gamma}_{\rm L} =
  \left( \begin{array}{cc}
      \Gamma(\varepsilon)/2 & 0 \\
      0 & 0
    \end{array}
  \right), \quad\boldgreek{\Gamma}_{\rm R} =
  \left( \begin{array}{cc}
      0 & 0 \\
      0 & \Gamma(\varepsilon)/2
    \end{array}
  \right)\;.
\end{gather*}
The matrix element $\Gamma(\varepsilon)$ is either constant in the case of WBL, or is taken to be a single Lorentzian
\cite{wime94},
\begin{equation}\label{eq:AppLWF}
  \Gamma( \varepsilon ) = \Gamma \frac{ W^2 }{ \varepsilon^2 + W^2 }\;.
\end{equation}
The latter is compatible with the general ansatz given in Eq.\ \eqref{eq:LWFexp} and
is chosen such that WBL is attained for $W\to\infty$.
For the time dependence of the bias voltage we take a rectangular pulse, i.e., 
\begin{equation}
  \label{eq:pulse}
  V_{\rm bias}(t) = \frac{V_{\rm max}}{2} \left[\tanh\left( \frac{t}{t_{\rm s}} \right)  
-\tanh\left( \frac{t-t_{\rm p} }{t_{\rm s}} \right)\right] \;,
\end{equation}
which is characterized by the pulse length $t_{\rm p}$.
The finite switching time $t_{\rm s}$ reflects the experimental
situation (e.g.\ as reported in Refs.\ \cite{fuau+03,fuha+06}). 
In the following calculations we use $t_{\rm s} = 1/\Gamma$ and 
$V_{\rm max} = 3\Gamma$.
The equations of motion obtained in Sect.\ \ref{sec:EOM} for the WBL and the LLWF, respectively, 
are propagated using a fourth-order Runge-Kutta scheme \cite{prfl+92} with constant time step $\delta t = 0.02/\Gamma$. 
We use $N_{\rm F} = 120$ auxiliary modes for all calculations.

\begin{figure}[tbp!]
  \includegraphics[scale=0.3]{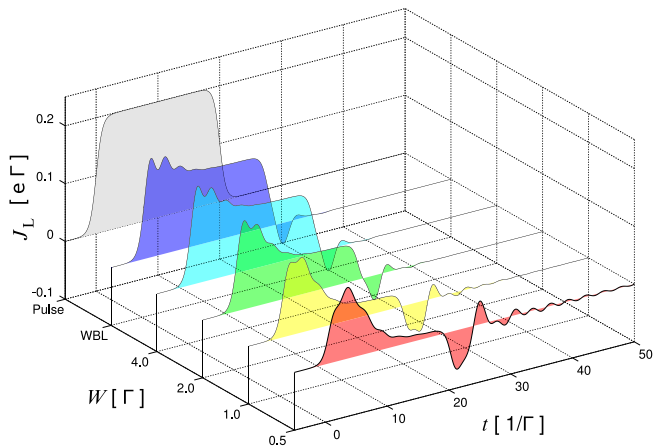}
  \caption{(color online) Time-resolved current through left barrier $J_{\rm L}$ for 
    different values of $W$ in Eq.\ \eqref{eq:AppLWF} driven by a
    bias voltage pulse according to Eq.\ \eqref{eq:pulse} with
    $V_{\rm bias}=3\Gamma$, $t_{\rm s}=1/\Gamma$ and  $t_{\rm p}=20/\Gamma$.
    The WBL corresponds to $W\to\infty$.
  }\label{fig:AppDQDTimeCurr}   
\end{figure}
\begin{figure}[b]
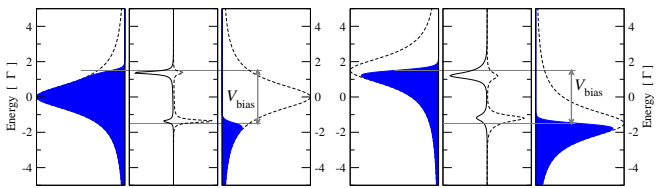

  \centering
  \includegraphics[scale=0.28]{Fig4a-DQD-Enscheme-Vb3-W1-NOSHIFT.eps}\hfill
  \includegraphics[scale=0.28]{Fig4b-DQD-Enscheme-Vb3-W1.eps}
  \caption{(color online) Energy scheme for $W=\Gamma$ at the plateau of the
    pulse with $V_{\rm bias} = V_{\rm max} = 3 \Gamma$.
    Outermost parts show
    $f_\alpha(\varepsilon)\Gamma(\varepsilon)$ for $\alpha={\rm
      L},{\rm R}$ (blue/dark-shaded areas).
    The inner part shows spectral densities of left and right levels.
    Left panel: $V_{\rm bias} = \mu_{\rm L} - \mu_{\rm R}$ and $\Delta_{\rm L} = \Delta_{\rm R} = 0$.
    Right panel: $V_{\rm bias} = \Delta_{\rm L} - \Delta_{\rm R}$ and $\mu_{\rm L} = \mu_{\rm R} = 0$.
  }\label{fig:AppDQDEnScheme}  
\end{figure}

%
Figure \ref{fig:AppDQDTimeCurr} shows the numerically obtained
current $J_{\rm L}$ as a function of time $t$ for different
widths $W$ in response to the same pulse of length $t_{\rm p} = 20/\Gamma$. The current shows a transient behavior at
the beginning and after the end of the pulse. For sufficiently long pulses it settles to a new stationary value
according to the plateau bias voltage $V_{\rm bias} = V_{\rm max}$. Notice that this situation for a structured reservoir 
is different from initially having $\mu_{\rm L} - \mu_{\rm R} = V_{\rm max}$. In the latter case the chemical potential and
the center of the level-width function [Eq.\ \eqref{eq:AppLWF}] are shifted with respect to each other. The two
distinct situations are illustrated in Fig.\ \ref{fig:AppDQDEnScheme}. We adopt the physical relevant situation 
shown in the right panel (see also Ref.\ \cite{jawi+94}).

\begin{figure}[t]
  \includegraphics[scale=0.6]{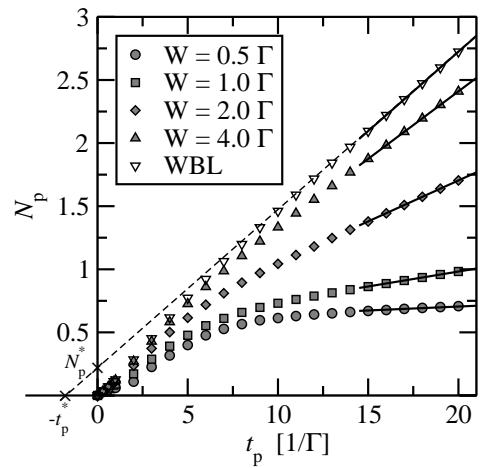}
  \caption{Number of pulse-induced tunneling-electrons $N_{\rm p}$ 
    vs pulse length $t_{\rm p}$. Symbols indicate numerical results for different values of $W$ in Eq.\ \eqref{eq:AppLWF}.
    Straight lines show result of linear fit in the respective
    range. The dashed line gives the two intercepts $N_{\rm
      p}^\star$ and $t_{\rm p}^\star$, respectively, for the case of the WBL.
  }\label{fig:AppDQDNpTp}   
\end{figure}

\begin{figure}[b]
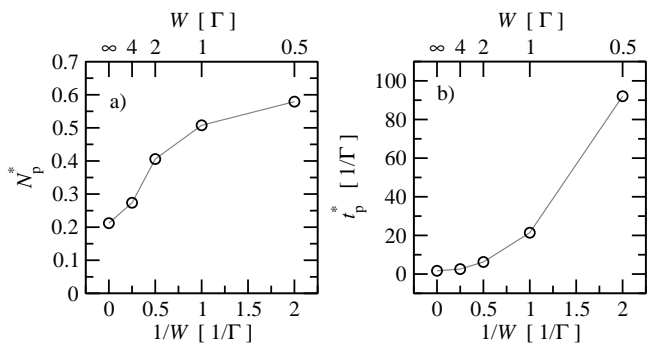

  \includegraphics[scale=0.49]{Fig6a-Np-Tp-Fit-1.eps}
  \includegraphics[scale=0.49]{Fig6b-Np-Tp-Fit-2.eps}
  \caption{Results of the linear fit to $N_{\rm p}(t_{\rm p})$ shown in Fig.\ \ref{fig:AppDQDNpTp}: 
    a) $N_{\rm p}$-intercept and b) $t_{\rm p}$-intercept.
  }\label{fig:AppDQDFit}   
\end{figure}

At any rate, the stationary current is found to vanish for $W\to 0$, which is an artifact of the level-width function given by
Eq.\ \eqref{eq:AppLWF}. The ringing behavior at the beginning and after the pulse is qualitatively similar for all
values of $W$. However, for small $W$ the damping of the current oscillations is weaker. In experiments the direct observation
of the ringing may be obscured by capacitive effects or the resolution of the ampere meter. Therefore, one considers 
the time-averaged or time-integrated current as a function of
pulse length \cite{wija+93,fuau+03}. The latter yields the number of
pulse-induced tunneling electrons,
\begin{equation}\label{eq:AppNpTp}
  N_{\rm p} (t_{\rm p}) = \int\limits^\infty_{-\infty } d t \left[ J (t) - J_0 \right]\;,
\end{equation}
where $J_0$ is the stationary current without pulse and $J(t) = J_{\rm L}(t)$.
In Fig.\ \ref{fig:AppDQDNpTp} we show $ N_{\rm p}$ as a function
of pulse length $t_{\rm p}$ for various
values of $W$. One observes an increase in the number of tunneling electrons with increasing pulse length.
Remembering the time dependence as shown in Fig.\ \ref{fig:AppDQDTimeCurr} it is clear that for short pulses
$N_{\rm p}$ is dominantly determined by the transient part of
the current. For sufficiently long pulses, however, the main
contribution comes from the new stationary current $J_{\rm stat}$ and one expects 
$N_{\rm p}  \propto t_{\rm p}$ with a slope given by $J_{\rm stat}$. This asymptotic behavior is shown in Fig.\ \ref{fig:AppDQDNpTp}
by the straight lines which have been obtained from a linear fit to the numerical data. The slope was fixed by
independently calculating $J_{\rm stat}$ using stationary NEGF formalism \cite{haja07}. The fitting procedure
yields the $N_{\rm p}$-intercept denoted by $N_{\rm p}^\star$ (cf.\
dashed line in Fig.\ \ref{fig:AppDQDNpTp}), which provides a measure of how transient the current response actually was. If the
current would instantaneously switch to the new stationary value one would get $N_{\rm p} =J_{\rm stat} t_{\rm p}$
and the $N_{\rm p}$-intercept would vanish. Non-vanishing values
of  $N_{\rm p}^\star$ reflect the additional transient contributions to the
current. Figure \ref{fig:AppDQDFit}a shows a stronger transient response for smaller values of $W$ which is in
accordance with the observations for the time-resolved current. 
In any case the net excess is positive since the transient response
following the switching-on outbalances the one after the switching-off.

Using the $N_{\rm p}$-intercept and the stationary current
one can also calculate the pulse length that would be necessary to yield the same number of tunneling
electrons if the DQD would switch instantaneously, $t_{\rm p}^\star = N_{\rm p}^\star/J_{\rm stat}$ (cf.\
dashed line in Fig.\ \ref{fig:AppDQDNpTp}). This quantity is
shown in Fig.\ \ref{fig:AppDQDFit}b.
It gives a measure for the pulse length at which transient and stationary
contributions are of similar size. Therefore, the transient
response for pulses with $t_{\rm p}\gg t_{\rm p}^\star$ becomes negligible. 

\section{Summary}
We have presented a propagation scheme for time-dependent
electron transport which is based on non-equilibrium Green
functions.
It relies on quantities with a single time argument
which allows for a straightforward numerical implementation with
standard differential equation solvers.

The basis of our scheme is a reformulation of the well-known expression
\cite{jawi+94} for the current $J(t)$ 
by means of the density matrix $\bg{\sigma}(t)$ and newly
introduced current matrices $\bg{\Pi}(t)$, cf.\ Eq.\ \eqref{eq:DefPi}. 
Decomposing these matrices into energy-resolved expressions allows
to obtain a closed set of coupled equations of motion for $\bg{\sigma}(t)$ and
$\bg{\Pi}(\varepsilon,t)$. Thereby, one has to consider another
energy-resolved quantity $\bg{\Omega}(\varepsilon,\varepsilon';t)$ given in Eq.\ \eqref{eq:DefOmega}.
The equations of motion are given by Eqs.\ \eqref{eq:EOMSEDM2}, \eqref{eq:EOMPi} and
\eqref{eq:EOMOmega}.

For a numerical implementation of these equations we propose using
an expansion of the Fermi function and a parameterization of the
level-width function by a set of Lorentzians \cite{meta99}.
The error made by truncating the expansion can be
reduced by applying a fast converging decomposition \cite{crsa09}.
The matrix equations to be solved are \eqref{eq:EOMSEDM1}, 
\eqref{eq:EOMPiLor} and \eqref{eq:EOMOmegaLor}, respectively.
In the often applied wide-band limit the set of equations
simplifies since $\bg{\Omega}$ can be found explicitly in terms
of the current matrices, cf.\ Eq.\ \eqref{eq:PiAuxWBLEOM}.

Finally, we have applied our scheme to two illustrative examples:
the randomly fluctuating energy level and the response of a DQD
to a voltage pulse. In both cases a non-trivial driving was involved.
For the DQD we demonstrated the influence of structured reservoirs on 
the transient current response. This transient contribution may be 
quantified by using the number of pulse-induced tunneling electrons 
[Eq.\ \eqref{eq:AppNpTp}] as a function of the pulse-length.
For the fluctuating energy level we showed a good agreement of our
numerical calculations with analytic results obtained for the
stationary current. Moreover, we found an enhancement of the current
due to the stochastic driving. The study of this effect in more complex
systems might lead to interesting new applications. 
In general, we expect our method to be a valuable tool for investigating 
time-resolved electron transport in nanoscale devices.

\begin{acknowledgments}
We thank Cenap Ates for his valuable comments during the preparation of
the manuscript.
\end{acknowledgments}

\appendix
\section{Expansions of Self-Energies} \label{sec:AppExp}
In order to perform the energy integration in Eqs.\
\eqref{eq:defself} we expand the Fermi function in terms of a 
finite sum over simple poles. This procedure yields the
expression given in Eq.\ \eqref{eq:ExpFermiFun}. 
The poles are given by $\chi_{\alpha p}^+=\mu_\alpha + x_p/\beta = \left( \chi_{\alpha p}^- \right)^*$.
Instead of using the Matsubara expansion \cite{ma90},
with poles $x_p=\imath\pi(2p{-}1)$,
we use a partial fraction decomposition of the 
Fermi function \cite{crsa09}, which converges much faster than
the standard Matsubara expansion.
Furthermore, it allows to estimate the error made by truncating
the sum [Eq.\ \eqref{eq:ExpFermiFun}] at $N_\mathrm{F}$ terms.
For this decomposition the poles $x_p=\pm 2\sqrt{z_p}$ are given by
the eigenvalues $z_p$ of the ${N_\mathrm{F}}{\times}{N_\mathrm{F}}$ matrix \cite{crsa09}
\begin{equation}
  Z_{ij} = 2i(2i{-}1)\delta_{j,i+1}-2{N_\mathrm{F}}(2{N_\mathrm{F}}{-}1)\delta_{i{N_\mathrm{F}}}\,.
\end{equation}
We take the branch of the root $\sqrt{z_p}$ such that 
$\Im(x_p)>0$ for all $p$. 
Thus all poles $\chi_p^+$ ($\chi_p^-$) are in the upper (lower)
complex plane.

Given the expansion [Eq.\ \eqref{eq:ExpFermiFun}] one can evaluate the energy integrals
by a contour integration in the upper or lower complex plane depending on the
sign of $t-t_1$. Thereby, the integration becomes a (finite) sum of the residues.

\section{Noise-Averaged Current for RLM} \label{sec:AppNoise}
The noise-averaged net-current, $\mean{J_{\rm net}(t)} = \mean{J_{\rm L}(t) - J_{\rm R}(t)}/2$, 
can be obtained from the general expression for the time-dependent current 
[Eq.\ \eqref{eq:GenCurrent1}], 
\begin{eqnarray}
  \mean{J_{\rm net}(t)} &=& e\, \Re \Tr \left\{ \int\limits_{-\infty}^{\infty} dt_1 
      \mean{ \mathbf{G}^\mathrm{r} (t, t_1)} \right.\notag\\
      &&\left.\quad\times\left[ \boldgreek{\Sigma}^<_{\rm L} (t_1,t) - \boldgreek{\Sigma}^<_{\rm R} (t_1,t) \right]
      \right\} \;, \label{eq:NAvgCurr}
\end{eqnarray}
where a symmetric coupling, $\boldgreek{\Gamma}_{\rm L} (\varepsilon, t_1, t) = \boldgreek{\Gamma}_{\rm R} (\varepsilon, t_1, t)$,
is assumed. For the resonant level model all quantities are scalars and in particular for the setting considered in
Sec.\ \ref{sec:RLMNoise} one has
\begin{equation}
  G^\mathrm{r} (t, t_1) = -\imath \Theta(t-t_1) \exp\left[ -\imath \int^t_{t_1} dt' \varepsilon_{\rm d} (t') - \frac{\Gamma}{2} (t-t_1)\right]\;.
\end{equation}
In order to evaluate Eq.\ \eqref{eq:NAvgCurr} we need 
the average of the fluctuating exponential function in $ G^\mathrm{r} (t, t_1)$
which is obtained by using the cumulant expansion, i.e.,
\begin{eqnarray}
  &&\mean{ \exp\left[ -\imath \int^t_{t_1} dt' \varepsilon_{\rm d} (t') \right] }\notag\\
  &&= \exp\left[ -\frac{1}{2} \int^t_{t_1} d\tau_1 \int^t_{t_1} d\tau_2 
    \mean{ \varepsilon_{\rm d} (\tau_1) \varepsilon_{\rm d} (\tau_2) } \right] \notag\\
  &&= \exp\left[ -\frac{1}{2} \int \frac{d\omega}{2\pi} c(\omega) 
    \left| \int^t_{t_1} d\tau e^{-\imath \omega \tau } \right|^2 \right] \notag \\
  &&= \exp\left[ -\frac{1}{2} \int \frac{d\omega}{2\pi} \frac{c(\omega) }{\omega^2}
    4 \sin^2 \left( \frac{\omega (t-t_1)}{2}  \right) \right] \;.
\end{eqnarray}
In the derivation we have used the properties of the noise [Eqs.\ \eqref{eq:NoiseProp}]
and introduced the Fourier transform of $c(\tau_1 - \tau_2)$ which is denoted by $c(\omega)$.

Thus, the noise-averaged retarded Green function does only
depend on the time difference 
and the time-averaged current is given by a Landauer-type expression \cite{jawi+94}
\begin{equation}\label{eq:TaNAvgCurr}
  \overline{ \mean{J_{\rm net}(t)} } = \frac{e \Gamma }{2} \int \frac{d \varepsilon}{2\pi}
  \left[ f_{\rm L}(\varepsilon) - f_{\rm R}(\varepsilon)\right] A(\varepsilon) \;,  
\end{equation}
where the spectral density $A( \varepsilon )$ is given by the time and noise averaged retarded 
Green function,
\begin{eqnarray}
  A( \varepsilon ) 
  &=& -2\, \Im \int\limits_{0}^{\infty} d\tau \mean{ G^\mathrm{r} (t, t- \tau) } e^{\imath \varepsilon \tau}\notag\\
  &=& \int\limits_{-\infty}^{\infty} d\tau e^{\imath \varepsilon \tau - \Gamma |\tau|/2}\\
  &&\qquad\exp\left[ -\frac{1}{2} \int \frac{d\omega}{2\pi} \frac{C(\omega) }{\omega^2}
    4 \sin^2 \left( \frac{\omega |\tau|}{2}  \right) \right] \;.\notag
\end{eqnarray}
For white noise one has $c(\omega) = \gamma$ and the
fluctuations lead to a trivial broadening 
of the spectral density.



\begin{thebibliography}{10}
\bibitem{fuha+06}
T. Fujisawa, T. Hayashi, and S. Sasaki, Rep. Prog. Phys. {\bf 69},  759
  (2006).
\bibitem{koel+06}
L.~P. Kouwenhoven, J.~M. Elzerman, R. Hanson, L.~H.~Willems van Beveren, and
  L.~M.~K. Vandersypen, phys. stat. sol. (b) {\bf 243},  3682  (2006).
\bibitem{fuau+03}
T. Fujisawa, D.~G. Austing, Y. Tokura, Y. Hirayama, and S. Tarucha, J. Phys.:
  Condens. Matter {\bf 15},  R\,1395  (2003).
\bibitem{hafu+03}
T. Hayashi, T. Fujisawa, H.~D. Cheong, Y.~H. Jeong, and Y. Hirayama, Phys. Rev.
  Lett. {\bf 91},  226804  (2003).
\bibitem{pejo+05}
J.~R. Petta, A.~C. Johnson, J.~M. Taylor, E.~A. Laird, A. Yacoby, M.~D. Lukin,
  C.~M. Marcus, M.~P. Hanson, and A.~C. Gossard, Science {\bf 309},  2180
  (2005).
\bibitem{kobu+06}
F.~H.~L. Koppens, C. Buizert, K.~J. Tielrooij, I.~T. Vink, K.~C. Nowack, T.
  Meunier, L.~P. Kouwenhoven, and L.~M.~K. Vandersypen, Nature {\bf 442},  766
  (2006).
\bibitem{jawi+94}
A.-P. Jauho, N.~S. Wingreen, and Y. Meir, Phys. Rev. B {\bf 50},  5528  (1994).
\bibitem{haja07}
H. Haug and A.-P. Jauho, {\em Quantum Kinetics in Transport and Optics of
  Semiconductors}, 2nd revised ed. (Springer, Berlin, 2007).
\bibitem{stwi96}
C.~A. Stafford and N.~S. Wingreen, Phys. Rev. Lett. {\bf 76},  1916  (1996).
\bibitem{cale+03}
S. Camalet, J. Lehmann, S. Kohler, and P. H\"{a}nggi, Phys. Rev. Lett. {\bf
  90},  210602  (2003).
\bibitem{ar05}
L. Arrachea, Phys. Rev. B {\bf 72},  125349  (2005).
\bibitem{zhma+05}
Y. Zhu, J. Maciejko, T. Ji, H. Guo, and J. Wang, Phys. Rev. B {\bf 71},  075317
   (2005).
\bibitem{mogu+07}
V. Moldoveanu, V. Gudmundsson, and A. Manolescu, Phys. Rev. B {\bf 76},  085330
   (2007).
\bibitem{kust+05}
S. Kurth, G. Stefanucci, C.-O. Almbladh, A. Rubio, and E.~K.~U. Gross, Phys.
  Rev. B {\bf 72},  035308  (2005).
\bibitem{stpe+08}
G. Stefanucci, E. Perfetto, and M. Cini, Phys. Rev. B {\bf 78},  075425
  (2008).
\bibitem{caco+71}
C. Caroli, R. Combescot, P. Nozieres, and D. Saint-James, J. Phys. C {\bf 4},
  916  (1971).
\bibitem{jizh+08}
J. Jin, X. Zheng, and Y. Yan, J. Chem. Phys. {\bf 128},  234703  (2008).
\bibitem{meta99}
C. Meier and D.~J. Tannor, J. Chem. Phys. {\bf 111},  3365  (1999).
\bibitem{wesc+06}
S. Welack, M. Schreiber, and U. Kleinekath\"ofer, J. Chem. Phys. {\bf 124},
  044712  (2006).
\bibitem{ma90}
G.~D. Mahan, {\em Many Particle Physics}, 2nd ed. (Plenum, New York, 1990).
\bibitem{ga96}
C.~W. Gardiner, {\em Handbook of Stochastic Methods: {F}or Physics, Chemistry
  and the Natural Sciences ({S}pringer Series in Synergetics)} (Springer,
  Berlin, 1996).
\bibitem{mitr94}
G.~N. Milshtein and M.~V. Tret'yakov, J. Stat. Phys. {\bf 77},  691  (1994).
\bibitem{gaha+98}
L. Gammaitoni, P. H\"{a}nggi, P. Jung, and F. Marchesoni, Rev. Mod. Phys. {\bf
  70},  223  (1998).
\bibitem{wime94}
N.~S. Wingreen and Y. Meir, Phys. Rev. B {\bf 49},  11040  (1994).
\bibitem{prfl+92}
W.~H. Press, B.~P. Flannery, S.~A. Teukolsky, and W.~T. Vetterling, {\em
  Numerical Recipes in {C}: {T}he Art of Scientific Computing}, 2nd ed.
  (Cambridge University Press, Cambridge, 1992), p.\ 994.
\bibitem{wija+93}
N.~S. Wingreen, A.-P. Jauho, and Y. Meir, Phys. Rev. B {\bf 48},  8487  (1993).
\bibitem{crsa09}
A. Croy and U. Saalmann, Phys. Rev. B {\bf 80},  073102  (2009).
\end{thebibliography}
\end{document}